\documentclass[aps,pra,superscriptaddress,12pt]{revtex4-1}

\usepackage{amsmath}
\usepackage{graphicx}
\usepackage{color}

\begin{document}
\title{Heralded Quantum Entanglement between Distant Matter Qubits}
\date{\today}
\author{Wen-Juan Yang}
\affiliation{State Key Laboratory of Low Dimensional Quantum Physics, Tsinghua University, Beijing 100084, People's Republic of China}
\affiliation{Synergetic Innovation Center of Quantum Information and Quantum Physics, University of Science and Technology of China, Hefei, Anhui 230026, China}
\author{Xiang-Bin Wang}
\email{xbwang@mail.tsinghua.edu.cn}
\affiliation{State Key Laboratory of Low Dimensional Quantum Physics, Tsinghua University, Beijing 100084, People's Republic of China}
\affiliation{Synergetic Innovation Center of Quantum Information and Quantum Physics, University of Science and Technology of China, Hefei, Anhui 230026, China}
\affiliation{Jinan Institute of Quantum Technology, Shandong
Academy of Information and Communication Technology, Jinan 250101,
People's Republic of China}
\begin{abstract}
We propose a scheme to realize heralded quantum entanglement between two distant matter qubits using two $\Lambda$ atom systems. Our proposal does not need any photon interference. We also present a general theory of outcome state of non-monochromatic incident light and finite interaction time.
\end{abstract}
\maketitle
Quantum entanglement is a key ingredient in the study of loophole free test of quantum non-locality \cite{nonlocality1, nonlocality2, nonlocality3} and also in quantum information processing \cite{communication1, duan, communication5, communication6, communication7,communication8, repeater}. For long distance quantum communication, quantum entanglement between stationary qubits is often needed\cite{entanglementswapping, duan}.

There are lots of schemes and experiments to create quantum entanglement between matter qubits \cite{duan, communication8, feng, singlephoton, singlephoton2, entanglement1,  entanglement2, entanglement3, entanglement4, entanglement5, entanglement6, mv, entanglement7}. Unheralded quantum entanglement has been demonstrated by several groups\cite{entanglement2, entanglement3, entanglement4, entanglement5, entanglement6, mv}. The inevitable photon loss including channel loss and detection loss can cause severe loopholes \cite{loophole} in experiments such as quantum non-locality test \cite{nonlocality11, nonlocality22, nonlocality33, nonlocality44} and secure quantum communications.
The photon loss means that unfair sampling is actually possible therefore the experimental result with significant photon loss for non-locality test is undermined. Moreover, in a quantum key distribution, Eve can attack users' detectors and pretends her action to be photon loss.
 All these loopholes can be resolved  by the heralding mechanism \cite{quantummemory} with matter qubits. A heralded quantum entanglement \cite{duan, singlephoton, singlephoton2, entanglement2, communication8, feng, entanglement1, entanglement7} announces at which time an entangled state is prepared successfully over channel loss. Since we only need to consider those heralded events therefore the channel loss can be actually disregarded. Moreover, if the entangled state is on matter qubits rather than photons, the detection efficiency is almost perfect \cite{duan}. Therefore, effectively generating heralded entangled state on matter qubits is the central issue in the study of fundamental quantum mechanics and long distance quantum communication. However, so far distant heralded quantum entanglement has never been realized because all the existing schemes and experiments encounter the technical challenging of distant photon interferences. For example the famous DLCZ protocol \cite{duan} uses atomic ensemble as local memory and the quantum entanglement is generated through single photon interference. Single photon interference \cite{singlephoton, singlephoton2, entanglement2} requires micrometer precision of optical paths. To overcome this drawback, quantum entanglement generation schemes through two-photon interference are proposed \cite{communication8, feng, entanglement1, entanglement7}. In practice two-photon interference over long distance in free space is still a very challenging task. In free space, distant two-photon interference suffers from direction fluctuation of photon beam induced by mechanical vibrating of photon sources and atomosphere turbulence \cite{pcz}. Therefore, the two light spots will be poorly overlapped and the interference quality is significantly decreased. Note that here we consider the issue of the wave-fronts, a long coherent length of the wave trains does not help.
Due to these highly technical challenging, so far the realized distance for two-photon interference in free space is rather limited. Zhang et al.\cite{zhang} demonstrated two-photon interference over a distance of 3 meters.
To our knowledge, the highest record of the distance of two-photon interference in free space is $220m$ with one side free space and the other side optical fiber, which was done by Yong et al.\cite{pcz} by using highly sophisticated technologies of APT(acquiring, pointing and tracking) and SMF(single mode fiber). Technically, it will be even more challenging if one tries to improve the experiment with both sides being free space. For a goal of long distance quantum communication in the magnitude of 100 kilometers, perhaps we need schemes without photon interference. Here we propose such a scheme to create
heralded quantum entanglement between distant matter qubits without any photon interference, neither single-photon interference nor two-photon interference.

In what follows, we shall first show our scheme in generating distant-matter-qubit quantum entanglement and then we make a detailed study for our outcome quantum entanglement state both analytically and numerically.
\section*{Results}
\begin{figure}
  \includegraphics[width=0.5\textwidth]{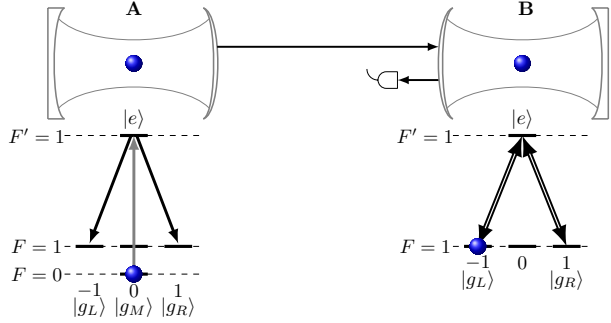}
  \caption{Schematic setup of our proposal. (A), A spontaneously decayed photon is emitted from cavity A and entangled with the two degenerate ground states of atom A. (B), The output photon from cavity A is scattered by cavity B with atom photon state swapped.}\label{model}
\end{figure}

\textbf{The scheme.} Schematic setup of our proposal is shown in Fig.~\ref{model}. There are two $\Lambda$ atoms and each atom is trapped in a cavity. We use two degenerate ground states of each atom for the encoding space. Initially, the atom trapped in cavity A in the ground state $|g_M\rangle$ is pumped by a short $\pi$ pulse to the excited state $|e\rangle$. Through the spontaneous decay process we have the entangled state of $\frac{1}{\sqrt{2}}(|g_L\rangle_A|L\rangle+|g_R\rangle_A|R\rangle)$ for the atom photon system. In our scheme, initially, atom in cavity B can be in any ground state as shown in Supplementary Material. Here for presentation simplicity, we set the initial state of the atom in cavity B to be $|\phi\rangle=|g_L\rangle_B$. The output photon of cavity A is the input signal of cavity B and the photon is then scattered by cavity B. According to Ref. \cite{infinity1}, after infinite time the state of the photon and atom in cavity B are swapped provided that the incident light is monochromatic. The final state of the whole system with two atoms and one photon is
\begin{eqnarray}
|\Psi\rangle&=&\frac{1}{\sqrt{2}}(|g_L\rangle_A|g_R\rangle_B-|g_R\rangle_A|g_L\rangle_B)
\otimes|R\rangle,\label{entangledstate}
\end{eqnarray}
This equation shows that if the detector clicks after infinitely long time then a high fidelity entangled state of atom A and B is created. In Ref. \cite{infinity1} Eq. (\ref{entangledstate}) is shown based on the assumption that the incident light is monochromatic and the atom photon interaction time is infinite. We now present a detailed study for the result based on the more actual situation of non-monochromatic incident light and the heralded event happens at finite time.

\textbf{Entanglement between Atom A and Photon.}
Here we consider a zero-temperature heat bath(a vacuum bath)  and initially, inside the cavity there is no photon while the atom is on excited state $|e\rangle$.
 For the ease of presentation, we have omit the atom-bath interaction at this moment and  we shall add this term later. Also, we present the derivation with complete Hamiltonian in Supplements I.
The Hamiltonian of our model is(set $\hbar=c=1$) \cite{quantumoptics}:
\begin{eqnarray}
H_A&=&H_{S_A}+H_{R}+H_{SR_A},\\
H_{S_A}&=&\omega_{e} |e\rangle_{AA}\langle e|+\sum_{i=L, R}\omega_{c} a_{i_A}^\dagger a_{i_A}+\sum_{i=L, R} (g_1 a_{i_A}|e\rangle_{AA}\langle g_i|+H.c.),\notag\\
H_{R}&=&\sum_{i=L, R}\int_{-\infty}^{\infty}\omega b_i^\dagger(\omega) b_i(\omega) d\omega ,\notag\\
H_{SR_A}&=&\sum_{i=L, R}\int_{-\infty}^{\infty}\sqrt{\frac{\kappa_1}{2\pi}}(-ib_i(\omega)a_{i_A}^{\dag}+H.c.)d\omega.\notag
\end{eqnarray}
Here $H_{S_A}$, $H_R$ and $H_{SR_A}$ stand for the Hamiltonian of the system A, the reservoir and the interaction between system A and reservoir, respectively. $\omega_{e}, \omega_{c}$ are the energy level of excited state of atom A and the resonant frequency of cavity A. $a_{i_A}(i=L, R)$, $|g_i\rangle_A(i=L, R)$ are the annihilation operator of the two polarization modes of photon and the corresponding ground state of atom A respectively. $g_1$ and $\kappa_1$ are the coupling strength of the CQED(cavity quantum electrodynamics) and the decay rate of the cavity A respectively.

Without loss of any generality, at any time $t$, the whole state of cavity A and its bath can be written in
\begin{eqnarray}
|\Phi(t)\rangle=|\tilde{\psi}(t)\rangle+ |\varphi(t)\rangle .
\end{eqnarray}
Here $|\tilde{\psi}(t)\rangle$ is an un-normalized state by which the photon number in the reservoir (i.e., outside the cavity) is zero. This means,
inside the cavity (the system), the state can be a linear superposition of excited atomic state with zero photon and ground atomic state with one photon. Moreover, $|\varphi(t)\rangle$ is an  un-normalized state by which the photon is in the reservoir (i.e. outside the cavity) and the atom inside the cavity can only be at the ground state. Note that we shall consider a continuous frequency mode for the reservoir, therefore $|\varphi(t)\rangle$ should have the form of $\int_{-\infty}^{\infty}
\varrho(\omega,t)|\phi(\omega,t)\rangle d\omega$ where $\varrho(\omega,t)$ is the amplitude functional on $\omega$ and $|\phi(\omega,t)\rangle$ is a state by which the photon is in the reservoir and the frequency is $\omega$ .

One can use the quantum trajectory method \cite{trajectory} to analyze the process. Moreover, following ref.\cite{smqo}, for our problem, we can reduce the method  to an effective Hamiltonian.
  Tracing over the subspace of reservoir, we can write the density operator of the system in the following form
\begin{eqnarray}
\hat{\rho}_{S_A}(t)=Tr_R[|\Phi(t)\rangle\langle\Phi(t)|]=\hat{\rho}_0(t)+\hat{\rho}_1(t),\label{rhosa}
\end{eqnarray}
where
\begin{eqnarray}
\hat{\rho}_0(t)&=&Tr_R \left(|\varphi(t)\rangle\langle\varphi(t)|\right),\\
\hat{\rho}_1(t)&=&Tr_R\left(|\tilde{\psi}(t)\rangle\langle\tilde{\psi}(t)|\right)= _R\langle 0 |\tilde{\psi}(t)\rangle\langle\tilde{\psi}(t)|0\rangle_R =|\tilde\psi_{in}(t)\rangle\langle \tilde \psi_{in}(t)|.
\end{eqnarray} Here $|0\rangle_R$ is the vacuum state for the reservoir
and $|\tilde \psi_{in}(t)\rangle $ has the form of
\begin{equation}
|\tilde \psi_{in}(t)\rangle =s_e(t)|e\rangle_A+s_1(t)|g_L\rangle_A|L\rangle_A+s_2(t)|g_R\rangle_A|R\rangle_A. \\ \label{state1}
\end{equation}
Given a vacuum bath initially, one can transform the equations above into the following master equation
\begin{eqnarray}
\dot{\hat{\rho}}_{S_A}(t)=(\mathcal{C}+\mathcal{D})\hat{\rho}_{S_A}(t),\label{dotrhosa}
\end{eqnarray}
where superoperators $\mathcal{C}$ and $\mathcal{D}$ are defined by
\begin{eqnarray}
\mathcal{C}\Omega &=& -i[H_{S_A},\Omega] -\sum_{i=L, R}{\kappa_1\over 2}\{a_{i_A}^\dagger a_{i_A},\Omega\},
\notag\\
\mathcal{D}\Omega &=&\sum_{i=L,R}\kappa_1(a_{i_A} \Omega a_{i_A}^\dagger),\notag
\end{eqnarray}
given any operator $\Omega$.
Substituting Eq. (\ref{rhosa}) into Eq. (\ref{dotrhosa}) we can get two separate equations
\begin{eqnarray}
\dot{\hat{\rho}}_0=\mathcal{D}\hat{\rho}_1\\
\dot{\hat{\rho}}_1=\mathcal{C}\hat{\rho}_1\label{heffa}
\end{eqnarray}
According to the definition of superoperators, we know that
\begin{equation}\label{scho}
\mathcal{C} \rho_1 = -i[H_{S_A},\rho_1] - \frac{\kappa_1}{2}\sum_{i=L, R}\{a_{i_A}^\dagger a_{i_A},\rho_1\}
\end{equation}
and $\mathcal{D}\rho_1=\sum\limits_{i=L, R}{\kappa_1\over 2}\{a_{i_A}\rho_1 a_{i_A}^{\dagger}\}$.
It is easy to see that Eq.(\ref{scho}) is equivalent to
\begin{eqnarray}\label{ee}
i{d\over dt}|\tilde{\psi}(t)\rangle_{in}=H_{eff_A}|\tilde{\psi}(t)\rangle_{in},
\end{eqnarray}
where the non-Hermitian effective Hamiltonian  $H_{eff_A}$ is
$
H_{eff_A} =H_{S_A} -i\sum\limits_{i=L, R}\frac{\kappa_1}{2}a_{i_A}^\dagger a_{i_A}.
$
In the above we have omitted the atom-bath interaction for simplicity. Obviously, one can add the atom-bath interaction in the same method with the following
\begin{eqnarray}
H_{eff_A}&=&\omega_{e} |e\rangle_{AA}\langle e|+\sum_{i=L, R}\omega_{c} a_{i_A}^\dagger a_{i_A}+\sum_{i=L, R} (g_1 a_{i_A}|e\rangle_{AA}\langle g_i|+ H.c.)
\notag\\
&&-i\frac{\gamma_1}{2}|e\rangle_{AA}\langle e|-i\sum_{i=L, R}\frac{\kappa_1}{2}a_{i_A}^\dagger a_{i_A}.\label{heff}
\end{eqnarray}
where $\gamma_1$ is the spontaneous decay rate. See supplementary Material for a detailed derivation. For clarity, we summarize the conclusion above by lemma 1:
\\{\bf Lemma 1} Given a vacuum bath initially, the intracavity initial state will evolve by the equation (\ref{ee}).\\
Through solving the Schr$\ddot{o}$dinger equation with the effective Hamiltonian in Eq.(\ref{heff}), we obtain
the time dependent amplitudes of Eq.(\ref{state1}) for the time evolution with the initial intracavity state $|e\rangle |0\rangle_{in}$ (zero photon with the atom being at the excited state):
\begin{eqnarray}
\begin{aligned}
s_e(t)&={{\mathrm{e}^{\nu t}}\over{\mu}}[\frac{i\Delta+{{\kappa_1}\over{2}}-{{\gamma_1}\over{2}}}{2}\sinh(\mu t)+\mu\cosh(\mu t)],\\
s_1(t)&=-{{ig_1}\over{\mu}} \mathrm{e}^{\nu t}\sinh(\mu t),\\
s_2(t)&=-{{ig_1}\over{\mu}} \mathrm{e}^{\nu t}\sinh(\mu t)
\end{aligned} \label{st}
\end{eqnarray}
where
 $\Delta=\omega_c-\omega_e$, $\mu=\sqrt{(\frac{i\Delta+{\kappa_1\over2}-{\gamma_1\over2}}{2})^2-2g_1^2}$ and $\nu=-\frac{i\Delta+{\kappa_1\over2}+{\gamma_1\over2}}{2}$. It is easy to see that these formulas satisfy the initial conditions $s_e(t=0)=1$ and $s_i(t=0)=0(i=1,2)$.

According to quantum regression formula \cite{smqo1}, the emission spectrum of left(right) circular photon along the cavity axis is \cite{smqo, smqo1}
\begin{eqnarray}
T(\omega)&=&{\kappa_1\over{2\pi}}\int_0^{\infty}dt\int_0^{\infty}dt^{\prime}e^{-i(\omega-\omega_c)(t-t^{\prime})}\langle a_{i_A}^\dagger(t)a_{i_A}(t^{\prime})\rangle\notag\\
&=&{\kappa_1\over{2\pi}}\int_0^{\infty}dt\int_0^{\infty}dt^{\prime}e^{-i(\omega-\omega_c)(t-t^{\prime})}s_i^{*}(t)s_i(t^{\prime})\notag\\&=&{\kappa_1\over{2\pi}}|\int_0^{\infty}dte^{i(\omega-\omega_c)t}s_i(t)|^2. \label{tw}
\end{eqnarray}
We assume $
\Delta=0$ and conditions of a bad cavity and a good emitter for cavity A $\gamma_1\ll\kappa_1$, $g_1\ll\kappa_1$(ensure $\mu$ being real, i.e. $g_1<\frac{\kappa_1-\gamma_1}{4\sqrt{2}}$). We find that the probability of photon emission from cavity A is:
\begin{equation}
p_{cav}=2\int_{-\infty}^{\infty}T(\omega)d\omega=\kappa_1\int_{0}^{\infty}(|s_1(t)|^2+|s_2(t)|^2)dt=\frac{\kappa_1 g_1^2}{2\nu(-\nu^2+\mu^2)}.
\end{equation}
This is in agreement with the weak coupling regime discussed in Ref\cite{emitter1, emitter2}.

The full width at half maximum (FWHM) of the emitted spectra is:
\begin{eqnarray}
\delta_{FWHM}=2\sqrt{-(\nu^2+\mu^2)+\sqrt{2(\nu^4+\mu^4)}}.
\end{eqnarray}

\begin{figure}
  \includegraphics[width=0.49\textwidth]{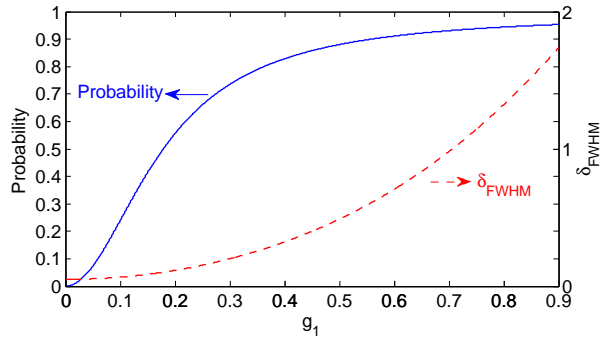}
  \caption{Probability(solid blue line)of photon emission from cavity A and the FWHM(dashed red line) of the emitted spectra versus the cavity coupling rate $g_1$ with $\kappa_1=5.0$ and $\gamma_1=0.05$(unit of x axis is $\kappa_1/5$).}\label{cavitya1}
\end{figure}

The probability(solid blue line) of photon emission from cavity A and the FWHM(dashed red line) of the emitted spectra versus the cavity coupling rate $g_1$ is shown in Fig. (\ref{cavitya1}) with $\kappa_1=5.0$ and $\gamma_1=0.05$. We can see that when $g_1$ rises, both $p_{cav}$ and $\delta_{FWHM}$ rise. The fidelity of the swapped state for cavity B is sensitive to the spectra width of incident photon, therefore we have to make a compromise of the parameter $g_1$ so that both the emission probability and the fidelity in swapping are satisfactorily good, say $g_1=\frac{(\kappa_1-\gamma_1)}{8\sqrt{2}}$ for example.

\begin{figure}
  \includegraphics[width=0.48\textwidth]{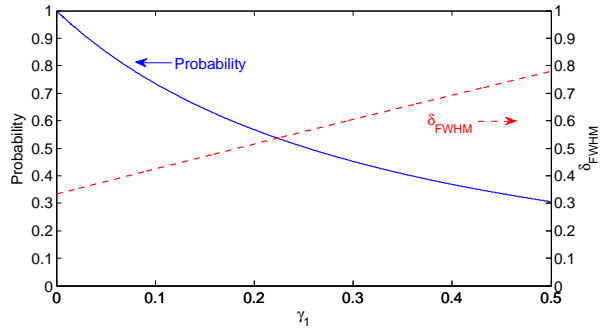}
  \caption{Probability(solid blue line)of photon emission from cavity A and the FWHM(dashed red line) of the emitted spectra versus the atomic decay rate $\gamma_1$ with $\kappa_1=5.0$ and $g_1=\frac{(\kappa_1-\gamma_1)}{8\sqrt{2}}$(unit of x axis is $\kappa_1/5$).}\label{cavitya2}
\end{figure}

Fig. (\ref{cavitya2}) shows the probability(solid blue line) of photon emission from cavity A and the FWHM(dashed red line) of the emitted photon spectra versus the atomic decay rate $\gamma_1$ with $\kappa_1=5.0$ and $g_1=\frac{(\kappa_1-\gamma_1)}{8\sqrt{2}}$. We can see that both the emission probability and the spectra width is sensitive to the atomic decay rate. As we shall show later, in order to obtain quantum entanglement of high quality between distant atoms one has to supress the decay rate of atom A.

We define $\tilde{s}_i(\omega)$ to be the normalized Fourier transform of $s_i(t)$,
\begin{eqnarray}
\tilde{s}_{i}(\omega)&=&\frac{1}{\sqrt{2\pi}}\frac{\int_{0}^{\infty}s_i(t)\mathrm{e}^{i(\omega-\omega_c)t}dt}{\int_{-\infty}^{\infty}|\tilde{s_i}(\omega)|^2d\omega}
=-\frac{2i\nu(\nu^2-\mu^2)}{\sqrt{2\pi\kappa_1}\mu g_1}(\frac{1}{i\delta_\omega+(\nu+\mu)}-\frac{1}{i\delta_\omega+(\nu-\mu)}),\label{somega}
\end{eqnarray}
where $i=1,2$ and $\delta_{\omega}=\omega-\omega_c$. Please note that the "$\delta_\omega$" here is different from that "$\Delta$" below Eq. (\ref{st}). From Eq. (\ref{tw}), we obtain that $T(\omega)\propto |\tilde{s}_i(\omega)|^2$. According to input-output theory \cite{quantumnoise}, the normalized final state of atom-photon system of cavity A is:
\begin{equation}
|\Phi(t=\infty)\rangle=|\varphi(t=\infty)\rangle=\sqrt{\frac{1}{2}}\int_{-\infty}^{\infty}e^{-i\omega t}s(\omega)(|g_L\rangle_A|L, \omega\rangle_{out}+|g_R\rangle_A|R, \omega\rangle_{out})d\omega,
\end{equation}
where $s(\omega)=\tilde{s}_{i}(\omega)$.

\textbf{State Swapping between Photon and Atom B.}
Consider the atom trapped in cavity B, a photon in a certain polarization state is injected into the cavity. We can formulate the time dependent evolution of the photon-atom state for cavity B through quantum trajectory theory. We shall use the input-output model and divide the process of photon scattering into two parts, i: direct reflection from the mirror outside the cavity, ii: first injected into the cavity and then leaking out.

Suppose initially the photon is inside the cavity B and the photon-atom state is $|\tilde{\varphi}(0)\rangle=|g_L\rangle_B\otimes\int_{-\infty}^{\infty}f(\omega)(\alpha|L, \omega\rangle+\beta|R, \omega\rangle)d\omega$. We can write time-dependent intracavity state of the photon-atom system in the form
\begin{eqnarray}
|\tilde{\varphi}(t)\rangle&=&c_e(t)|e\rangle_B+c_1(t)|g_L\rangle_B|L\rangle_B+c_2(t)|g_L\rangle_B|R\rangle_B\notag\\
&&+c_3(t)|g_R\rangle_B|L\rangle_B+c_4(t)|g_R\rangle_B|R\rangle_B.
\label{state}
\end{eqnarray}
We assume here that there is only one cavity mode resonant with the photon and the photon frequency width is far from the adjacent resonant modes of the cavity. Define
$c_j(t)=\sqrt{{1\over2\pi}}\int_{-\infty}^{\infty}f(\omega)\widetilde{c}_{j,\omega}(t)\mathrm{e}^{-i\omega t}d\omega$. The effective Hamiltonian is
\begin{eqnarray}
H_{eff_B}&=&\omega_{e} |e\rangle_{BB}\langle e|+\sum_{i=L, R}\omega_{c} a_{i_B}^\dagger a_{i_B}+\sum_{i=L, R} (g_2 a_{i_B}|e\rangle_{BB}\langle g_i|+ H.c.)\notag\\
&&-i\frac{\gamma_2}{2}|e\rangle_{BB}\langle e|-i\sum_{i=L, R}\frac{\kappa_2}{2}a_{i_B}^\dagger a_{i_B}.
\end{eqnarray}Solving
Schr$\ddot{o}$dinger equation $i\frac{\partial}{\partial t}{|\tilde{\varphi}(t)\rangle}=H_{eff_B}|\tilde{\varphi}(t)\rangle$ we obtain
\begin{eqnarray}
 \widetilde{c}_{e,\omega}(t)|e, \omega\rangle_B&=&- i\alpha \frac{g_2}{\eta }\mathrm{e}^{\lambda t}\sinh\eta t|e, \omega\rangle_B ,\notag\\
 \widetilde{c}_{1,\omega}(t)|g_L\rangle_B|L,\omega\rangle_B&=&\alpha(\frac{{2\rho -\lambda }}{{2\eta }}\mathrm{e}^{\lambda t}\sinh\eta t+\frac{1}{2}\mathrm{e}^{\lambda t}\cosh\eta t+\frac{1}{2}\mathrm{e}^{2\rho t})|g_L\rangle_B|L,\omega\rangle_B,\notag\\
 \widetilde{c}_{2,\omega}(t)|g_L\rangle_B|R,\omega\rangle_B&=&\beta \mathrm{e}^{2\rho t}|g_L\rangle_B|R,\omega\rangle_B,\notag\\
 \widetilde{c}_{3,\omega}(t)|g_R\rangle_B|L,\omega\rangle_B&=&0 ,\notag\\
 \widetilde{c}_{4,\omega}(t)|g_R\rangle_B|R,\omega\rangle_B&=&\alpha(\frac{{2\rho-\lambda }}{{2\eta }}\mathrm{e}^{\lambda t}\sinh\eta t + \frac{1}{2}\mathrm{e}^{\lambda t}\cosh\eta t\notag-\frac{1}{2}\mathrm{e}^{2\rho t})|g_R\rangle_B|R,\omega\rangle_B, \notag\\
\end{eqnarray}
where $\Delta=\omega_c-\omega_e$, $\delta_\omega=\omega-\omega_c$, $\lambda=-\frac{{-i\Delta-2i\delta_\omega+\frac{\kappa_2}{2}+\frac{\gamma_2}{2}}}{2}$, $\eta=\sqrt{(\frac{{-i\Delta-\frac{\kappa_2}{2}+\frac{\gamma_2}{2}}}{2})^2-2g_2^2 }$, $\rho=\frac{{i\delta_\omega-\frac{\kappa_2}{2}}}{2}$.

Now we consider a more exact model, the input-output model.  The photon as a wave train is initially outside the cavity B. As shown in Supplementary Material, the amplitude for photon inside cavity B should be the integration of time as
\begin{equation}
C_{i,\omega}(t)|\chi\rangle_i=-\sqrt{{\kappa_2}}\int_0^t\widetilde{c}_{i,\omega}(t')dt'|\chi\rangle_i,
\end{equation}
where $i=1,2,3,4$ and $|\chi\rangle_i$($i=1,2,3,4$) are $|g_L\rangle_B|L, \omega\rangle_B$, $|g_L\rangle_B|R, \omega\rangle_B$, $|g_R\rangle_B|L, \omega\rangle_B$ and $|g_R\rangle_B|R, \omega\rangle_B$ respectively. The explicit expression is shown in Supplementary Material.

 The wave train outside the cavity contains two parts. One is the amplitude that is directly reflected, the other is the beam that transmits inside the cavity and then leaks out. The interference of the two parts should be considered. When a photon is heralded, wave function of component $\omega$ should be:
\begin{eqnarray}
|\psi_\omega(t)\rangle&=&\sqrt{\kappa_2}(C_{1,\omega}(t)|g_L\rangle_B|L, \omega\rangle+C_{2,\omega}(t)|g_L\rangle_B|R, \omega\rangle
+C_{3,\omega}(t)|g_R\rangle_B|L, \omega\rangle
\notag\\&+&C_{4,\omega}(t)|g_R\rangle_B|R, \omega\rangle)+|g_L\rangle_B(\alpha|L,\omega\rangle+\beta|R,\omega\rangle)\label{modelm}.
\end{eqnarray}
Here the first term on the right side of the equation is due to the atom scattering process. The second term is due to the direct reflection from the left mirror. If the detection very late, one do not need to consider the interference between output photon from cavity B and direct reflection from cavity B.

With the spectral amplitude $f(\omega)$, we give wave function of the non-monochromatic incident case:
 \begin{equation}|\psi(t)\rangle=\mathrm{e}^{-i\omega_ct}\sqrt{{1\over2\pi}}\int_{-\infty}^{\infty}f(\omega)|\psi_\omega(t)\rangle \mathrm{e}^{-i\delta_\omega t}d\omega\label{wf}\end{equation} in schr$\ddot{o}$dinger picture. In particular, setting $\Delta=0$, $\delta_\omega=0$ and $\gamma_2=0$(the monochromatic incident case), we have,
\begin{equation}
|\psi(t=\infty)\rangle=(-\beta |g_L\rangle_B+\alpha |g_R\rangle_B)\otimes|R\rangle.
\label{stateswapping}
\end{equation}
This is in agreement with the existing results \cite{zoller, infinity1, infinity2, infinity3, infinity4}.

\textbf{Entanglement between Atoms.} In our proposed setup, the incident photon of cavity B is initially entangled with atom A. In such a case, according to Eq. (\ref{wf}) the tripartite state is
\begin{eqnarray}
|\psi(t)\rangle&=&cc_1(t)|g_L\rangle_A|g_L\rangle_B|L\rangle+cc_2(t)|g_R\rangle_A|g_L\rangle_B|R\rangle
+cc_3(t)|g_L\rangle_A|g_R\rangle_B|R\rangle,\label{stateswappingf}
\end{eqnarray}
where $\Delta=0$, $f(\omega)=s(\omega)$, $\alpha=\beta={1\over{\sqrt{2}}}$ and the explicit expression is shown in Supplementary Material.

\begin{figure}
  \includegraphics[width=0.48\textwidth]{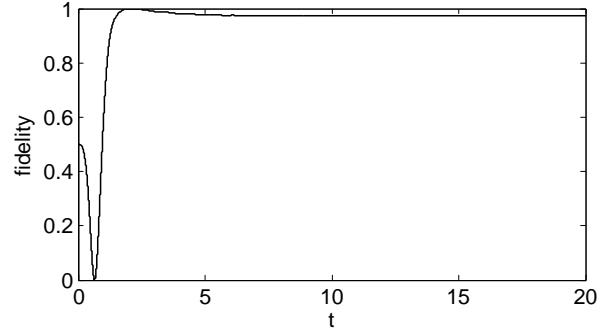}
  \caption{Fidelity of outcome entangled state versus time $t$. Here we set $\kappa_1=5.0$, $\gamma_1=0.5$, $g_1=\frac{(\kappa_1-\gamma_1)}{8\sqrt{2}}$, $\kappa_2=2.0$, $\gamma_2=1.0$ and $g_2=5.0$(unit of x axis is $2/\kappa_2$).}\label{cavitybft}
\end{figure}

\begin{figure}
  \includegraphics[width=0.49\textwidth]{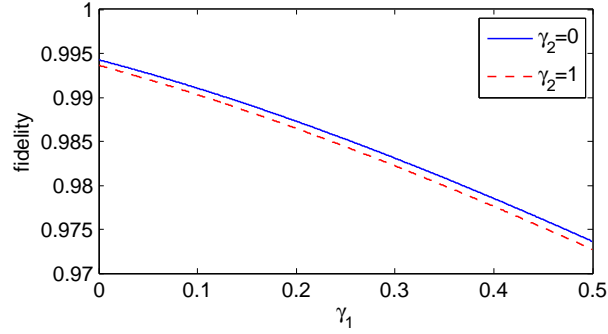}
  \caption{Fidelity of the outcome entangled state versus the decay rate $\gamma_1$ of atom A with different values of decay rates of atom B, $\gamma_2=0$ and $\gamma_2=1$. Here we have set $\kappa_1=5.0$, $g_1=\frac{(\kappa_1-\gamma_1)}{8\sqrt{2}}$, $\kappa_2=2.0$ and $g_2=5.0$(unit of x axis is $\kappa_2/2$).}\label{cavitybf}
\end{figure}

\begin{figure}
  \includegraphics[width=0.49\textwidth]{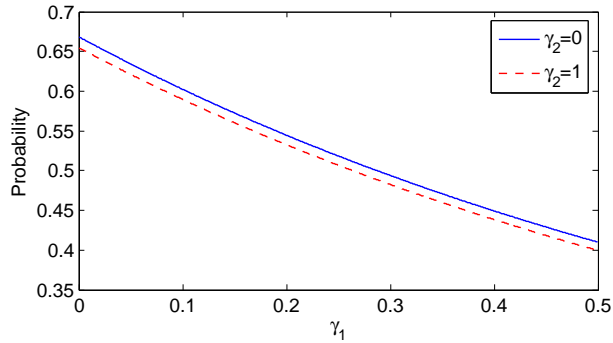}
  \caption{Probability of photon emission versus the decay rate $\gamma_1$ of atom A with different values of decay rates of atom B, $\gamma_2=0$ and $\gamma_2=1$. Here we have set  $\kappa_1=5.0$, $g_1=\frac{(\kappa_1-\gamma_1)}{8\sqrt{2}}$, $\kappa_2=2.0$ and $g_2=5.0$(unit of x axis is $\kappa_2/2$).}\label{cavityb2}
\end{figure}

The fidelity $f=|\langle\Psi|\psi(t)\rangle|^2$ for the outcome entangled state versus time $t$ is shown in Fig. (\ref{cavitybft}). Here $|\Psi\rangle$ and $|\psi(t)\rangle$ are defined in Eq. (\ref{stateswappingf}) and Eq. (\ref{entangledstate}) respectively. We can see that a high fidelity outcome entangled state can be obtained at finite time rather than an infinite time as requested by prior art theory\cite{zoller, infinity1, infinity2, infinity3, infinity4}. We choose $t_{start}=2.05/({\kappa_2\over 2})$ as the time point after which the outcome entangled state is good enough. Note that here should add a term of travel time $L/c$ from cavity A to B where $L$ is the distance between cavity A and B and $c$ is the light speed. The travel time will only cause a uniform time translation for the whole system, it does not change the time interval of each individual cycle from pumping to heralding, hence it does not change the system repetition rate. In Fig.
(\ref{cavitybf}) we plot the fidelity $f$ of the outcome entangled state versus the decay rates of atoms. We can see that the fidelity decreases
apparently with $\gamma_1$. This is because the spectra width $\delta_{FWHM}$ increases fast with $\gamma_1$, as shown in Fig. (\ref{cavitya2}).
The photon emission probability of cavity B is shown in Fig. (\ref{cavityb2}). The parameters are the same as those in Fig. (\ref{cavitybf}). After the time point $t_{start}$ a waiting time interval of $\Delta t_{wait}=14.95/({\kappa_2\over 2})$ is sufficient for a large probability of heralding. Even though we wait for longer, the probability of obtaining a heralded event hardly changes. One can realize such entangled state in many real atoms. For example, the $D_2$ line($5^2S_{1/2} \rightarrow 5^2P_{3/2}$)of
 $^{87}Rb$ atom  which has been used in the experiment already\cite{entanglement1}. The atom states $|F=1, m_F=-1\rangle$, $|F=1, m_F=1\rangle$ and $|F^{'}=1, m_{F^{'}}=0\rangle$ correspond to $|g_L\rangle$,
 $|g_R\rangle$ and $|e\rangle$ respectively.  The parameters of this system can set to be $(g_1, \kappa_1, \gamma_1)/2\pi=(1.2, 15, 1.5)Mhz$ and $(g_2, \kappa_2, \gamma_2)/2\pi=(15, 6, 3)Mhz$. This parameter setting needs $t_{start}=0.11\mu s$ and $\Delta t_{wait}=0.79\mu s$. The outcome state fidelity $f=0.9727$ and
 overall heralding probability $P=p_{cav}\times p=12.21\%$.

\section*{Discussion}
We have proposed a scheme to generate heralded quantum entanglement between two distant matter qubits. In our proposal neither single photon interference nor two-photon interference are involved. In addition, we have presented an analytical solution of the atom-photon entanglement and state swapping in CQED. With some specific parameter settings a high fidelity matter-qubit entanglement can be created.

\textbf{Acknowledgements}

We acknowledge professor B. Zhao, C. Z. Peng, Q. Zhang and Y. Cao at USTC, Q. Wang at NUPT, S. C. Wang and T. Chen for helpful discussion. This work was supported in part by the National High-Tech Program of
China grant No. 2011AA010800 and 2011AA010803, NSFC grant No.11174177, 60725416 and 11474182.

\textbf{Author Contributions}

X. B. W. proposed this work, W. J. Y. did the calculations and drew the figures. W. J. Y. and X. B. W. wrote the manuscript.

\textbf{Additional Information}

\texttt{Supplementary information} accompanies this paper at http://www.nature.com/\\
scientificreports

\texttt{Competing financial interests:} The authors declare no competing financial interests.

\newpage
\newpage
\end{document}